\documentclass[aps,nofootinbib,preprint]{revtex4}
\usepackage{graphicx}
\usepackage{amsfonts}

\def\Lie{\pounds}

\begin{document}

\title{Holographic Duals of Black Holes in Five-dimensional Minimal Supergravity}

\author{Chiang-Mei Chen} \email{cmchen@phy.ncu.edu.tw}
\affiliation{Department of Physics and Center for Mathematics and Theoretical Physics, National Central University, Chungli 320, Taiwan}
\affiliation{AEI, Max-Planck-Institut f\"{u}r Gravitationsphysik, Am M\"{u}hlenberg 1, D-14476 Potsdam, Germany}

\author{John E. Wang} \email{jwang@niagara.edu}
\affiliation{Department of Physics, Niagara University, Niagara University, NY 14109-2044}
\affiliation{HEPCOS, Department of Physics, SUNY at Buffalo, Buffalo, NY 14260-1500}

\date{\today}

%% REVTEX4
%\maketitle

\begin{abstract}
We examine the dual conformal field theory for extremal charged black holes in five-dimensional minimal supergravity with 2 independent angular momenta.  The conformal field theory Virasoro algebra, central charge, and temperature are calculated.  Additionally the conformal field theory entropy is calculated using the Cardy formula and agrees with the Bekenstein-Hawking black hole entropy.  The central charges are directly proportional to the angular momentum components of the black hole. In five and higher dimensions, rotations of the spacetime correspond to rotations of the central charges leading to an apparent symmetry relating the conformal field theories dual to each black hole.  A rotationally invariant central charge, which is proportional to the total angular momentum, is used to discuss the supersymmetric BMPV black hole limits.
\end{abstract}

%% REVTEX4
%\pacs{empty}

\maketitle
%%%%%%%%%%%%%%%%%%%%%%%%%%%%%%%%%%%%%%%%%%%%%%%%%%%%%%%%%%%%%%%%%%%%%%
\section{Introduction}
%%%%%%%%%%%%%%%%%%%%%%%%%%%%%%%%%%%%%%%%%%%%%%%%%%%%%%%%%%%%%%%%%%%%%%

Black holes and their generalizatons have been an active source of research due to their mysterious property that their horizons obey a form of thermodynamics.  To understand this thermodynamic property, many approaches have been taken including Ref.~\cite{Guica:2008mu} which recently extended the Brown-Henneaux methods \cite{Brown:1986nw} to extremal Kerr black holes in four dimensions.  There have been several following works discussing various types of rotating black holes \cite{Lu:2008jk, Azeyanagi:2008kb, Hartman:2008pb, Chow:2008dp, Isono:2008kx, Azeyanagi:2008dk}.  This method involves the study of the boundary conditions for the metric and gauge fields for black holes.  A set of Virasoro generators is found which preserves these boundary conditions, and the  associated central charge and the temperature are calculated.\footnote{Earlier work in Refs.~\cite{Carlip:1998wz,Solodukhin:1998tc, Park:2001zn} applied Brown-Henneaux methods to non-extremal black holes.  These papers took an alternate approach by setting boundary conditions on the black hole horizon.}  Finding the central charge is an important step due to the Cardy formula, which states that a conformal field theory is governed by its central charge at high energies.  The main result of this method is that there is a two dimensional CFT dual holographic description of the black hole spacetime.  Further evidence of this duality is that the Bekenstein-Hawking entropy of the black hole is equal to the conformal field theory entropy as calculated using the Cardy formula.

In this paper we examine a class \cite{Chong:2005hr} of non-supersymmetric but extremal rotating black holes in five dimensions.  These charged black holes, which are solutions of  minimal supergravity, have two independent angular momenta and degenerate horizons.  These black holes reduce as a special case to the Breckenridge-Myers-Peet-Vafa (BMPV) solutions \cite{Breckenridge:1996is}.  Previously BMPV black holes have also been studied using string theory methods so it would be interesting to compare the previous string theory insights to the Brown-Henneaux approach.

According to Ref.~\cite{Guica:2008mu} we examine the near horizon limit, find the Virasoro generators and associated central charge, and as a check we compare the Bekenstein-Hawking black hole entropy to the Cardy formula. The central charges, temperature and entropy are
\begin{eqnarray}
c^\phi &=& \frac{3 \pi}{2 G_5 \hbar} \left[ a r_0^2 + b (2 \mu - a^2) \right] = \frac{6}{\hbar} J^\psi ,
\\
c^\psi &=& \frac{3 \pi}{2 G_5 \hbar} \left[ b r_0^2 + a (2 \mu - b^2) \right] = \frac{6}{\hbar} J^\phi ,
\\
T_{FT}^\phi &=& \frac{(r_0^2 + a^2) (r_0^2 + b^2) + a b q}{\pi r_0 [a r_0^2 + b (2 \mu - a^2)]} ,
\\
T_{FT}^\psi &=& \frac{(r_0^2 + a^2) (r_0^2 + b^2) + a b q}{\pi r_0 [b r_0^2 + a (2 \mu - b^2)]} ,
\end{eqnarray}
\begin{equation}
S_{CFT} = \frac{\pi^2}3 c^\phi \, T_{FT}^\phi = \frac{\pi^2}3 c^\psi \, T_{FT}^\psi = \frac{\pi^2 [(r_0^2 + a^2) (r_0^2 + b^2) + a b q]}{2 G_5 \hbar r_0} = \frac{A}{4 G_5 \hbar} = S_{BH}  \ .
\end{equation}
As these solutions are five dimensional there are two independent angular momenta corresponding to angular momentum along two orthogonal spatial planes.  Two central charges and two temperatures can be calculated for each of the angular directions  $\phi,\psi$. In five dimensions there exists a symmetry for these space times corresponding to a rotation of the two planes.  This rotation according to the prescribed methods, changes the central charges and temperatures of the conformal field theory.  Although this suggests that there is a rotational symmetry relating the conformal field theories,  the central charges are related to the components of the angular momentum.  Rotating the overall direction of the angular momentum should have no physical effect on the gravitational solution, so the dual conformal field theory description should be invariant under rotations as well.  Searching for an invariant description, a rotationally invariant central charge, $C$, for the holographic theory is defined
\begin{equation}
C^2 = (c^\phi)^2 + (c^\psi)^2 \propto J^2 \ .
\end{equation}
As an application of this invariant central charge (angular momentum), we discuss how it can be used to find the supersymmetric BMPV limit where the angular momentum $J^\phi = -J^\psi$.

%%%%%%%%%%%%%%%%%%%%%%%%%%%%%%%%%%%%%%%%%%%%%%%%%%%%%%%%%%%%%%%%%%%%%%
\section{Review of the Brown-Henneaux approach}
%%%%%%%%%%%%%%%%%%%%%%%%%%%%%%%%%%%%%%%%%%%%%%%%%%%%%%%%%%%%%%%%%%%%%%

In this section, we briefly review the Brown-Henneaux \cite{Brown:1986nw} approach to examining black holes and finding their dual holographic conformal field theory descriptions.  In a gravitational theory the aim is to find a representation of the conformal symmetry including the Virasoro algebra, the central charge and the temperature.  Whenever these are found, the states of the black hole form a representation of a conformal field theory or there is a dual conformal field theory description of the black hole.

As we review below the Virasoro algebra of the conformal field theory arises as an asymptotic symmetry group for the near horizon limit of rotating black holes.  The central charges arises as an integral over the asymptotic boundary for the near horizon geometry, and the temperature of the CFT is the Frolov-Thorne temperature of the black hole horizon.

\subsection{Asymptotic Symmetric Group and Virasoro generators}
%%%%%%%%%%%%%%%%%%%%%%%%%%%%%%%%%%%%%%%%%%%%%%%%%%%%%%%%%%%%%%%%%%%%%%
For a gravitational theory with a gauge field, the asymptotic symmetries of any charged black hole solution includes the diffeomorphisms $\zeta$ which act on the metric and the gauge field
\begin{equation}
h_{\mu\nu} = \delta_\zeta g_{\mu\nu} = \Lie_\zeta g_{\mu\nu}, \qquad a_\alpha = \delta_\zeta A_\alpha = \Lie_\zeta A_\alpha,
\end{equation}
as well as the gauge transformations $\Lambda$
\begin{equation}
\delta_\Lambda A = d\Lambda.
\end{equation}
For spacetimes with a $U(1)$ symmetry it is possible to look for a Virasoro symmetry among the diffeomorphisms.  To do this one needs to impose asymptotic boundary conditions on the metric and gauge field given in for example \cite{Guica:2008mu, Hartman:2008pb}.   These asymptotic boundary conditions  admit the following diffeomorphism generators
\begin{equation}
\zeta_\epsilon^{\varphi^a} = - r \epsilon'(\varphi^a) \partial_r + \epsilon(\varphi^a) \partial_{\varphi^a},
\end{equation}
here $\varphi^a$ are angular coordinates for rotations.
In this paper $\varphi^a$ will be either $(\phi, \psi)$, or $(\phi, y)$ for Kaluza-Klein theory.  It can be checked that this generator forms the Virasoro algebra for a conformal field theory by taking the mode expansion  $\epsilon_n(\varphi^a) = - \mathrm{e}^{- i n \varphi^a}$.  The modes of the diffeomorphism generators are
\begin{equation}
\zeta_n^{\varphi^a} = - i n r \mathrm{e}^{- i n \varphi^a} \partial_r - \mathrm{e}^{- i n \varphi^a} \partial_{\varphi^a},
\end{equation}
which satisfy the Virasoro algebra without central charge.

\subsection{Central charge}
%%%%%%%%%%%%%%%%%%%%%%%%%%%%%%%%%%%%%%%%%%%%%%%%%%%%%%%%%%%%%%%%%%%%%%
The associated charge $\delta Q_{\zeta, \Lambda}$ with respect to the combined transformation $(\zeta, \Lambda)$ is defined by
\begin{equation} \label{Charge}
\delta Q_{\zeta, \Lambda} = \frac1{8 \pi G} \int_{\partial \Sigma} \left( k^g_\zeta[h; g] + k^A_{\zeta, \Lambda}[h, a; g, A] \right),
\end{equation}
where the contributions from the metric and gauge field are respectively \cite{Barnich:2001jy, Barnich:2005kq, Barnich:2007bf, Guica:2008mu, Hartman:2008pb}
\begin{eqnarray}
k^g_\zeta[h; g] &=& - \frac14 \epsilon_{\alpha \cdots \beta\mu\nu} \Bigl[ \zeta^\nu D^\mu h - \zeta^\nu D_\sigma h^{\mu\sigma} + \zeta_\sigma D^\nu h^{\mu\sigma} + \frac12 h D^\nu \zeta^\mu - h^{\nu\sigma} D_\sigma \zeta^\mu
\nonumber\\
&& \qquad + \frac12 h^{\sigma\nu} (D^\mu \zeta_\sigma + D_\sigma \zeta^\mu) \Bigr] dx^\alpha \wedge \cdots \wedge dx^\beta,
\\
k^A_{\zeta, \Lambda}[h, a; g, A] &=& \frac14 \epsilon_{\alpha \cdots \beta\mu\nu} \Biggl[ \left( 2 F^{\mu\sigma} h_\sigma{}^\nu - \frac12 h F^{\mu\nu}  - \delta F^{\mu\nu} \right) (\zeta^\rho A_\rho + \Lambda) - F^{\mu\nu} \zeta^\rho a_\rho - 2 F^{\rho\mu} \zeta^\nu a_\rho
\nonumber\\
&& \qquad + g^{\mu\rho} g^{\nu\sigma} a_\rho (\Lie_\zeta A_\sigma + \partial_\sigma \Lambda) \Biggr] dx^\alpha \wedge \cdots \wedge dx^\beta.
\end{eqnarray}
In the above expressions $h_{\mu\nu} = \delta_\zeta g_{\mu\nu} = \Lie_\zeta g_{\mu\nu}, \; a_\alpha = \delta_{\zeta,\Lambda} A_\alpha = \Lie_\zeta A_\alpha + d\Lambda$ and the operations of covariant derivatives, raising/lowering indices are computed with respect to the background metric $g_{\mu\nu}$.

The generators $\delta Q_m$ of (\ref{Charge}) include the previous diffeomorphisms but also boundary terms.  These charges can be identified \cite{Guica:2008mu} with the Virasoro generators $L_m$ except the Virasoro algebra now there has a central charge $Q^{\varphi^a}_{mn}$
\begin{equation} \label{Cterm}
Q^{\varphi^a}_{mn} = \frac1{8 \pi G} \int_{\partial \Sigma} k^g_{\zeta_m^{\varphi^a}} [\Lie_{\zeta_n^{\varphi^a}} g; g].
\end{equation}
The integral will be of the general form
\begin{equation} \label{Ccharge}
Q^{\varphi^a}_{mn} = - i \frac{\hbar}{12} (m^3 + \alpha m) \delta_{m+n,0} \, c^{\varphi^a},
\end{equation}
from which the central charge, $c^{\varphi^a}$, can be found.  The value of $\alpha$ is not essential since it can be changed by c-number shifts of the Virasoro generators of $L_m$.

The gauge transformation is chosen to compensate the diffeomorphism transformations in order to ensure the desired boundary condition for $a_\alpha$, which generically eliminates the gauge field contribution to the charge (\ref{Charge}).  In this paper, we will hence only focus on the central charge contributions from gravity.

\subsection{Temperature}
%%%%%%%%%%%%%%%%%%%%%%%%%%%%%%%%%%%%%%%%%%%%%%%%%%%%%%%%%%%%%%%%%%%%%%
The temperature of the conformal field theory, or the Frolov-Thorne temperature, can be obtained from considering the expansion of the quantum fields in the eigenmodes of the asymptotic energy $\omega$ and angular momenta $m_a$
\begin{equation}
\Phi = \sum \phi_{\omega m_a l} \mathrm{e}^{- i \omega \hat{t} + i m_a \hat{\varphi}^a} f_l (r, \theta).
\end{equation}
Generically we need the following transformation in order to obtain the near-horizon limits
\begin{equation}
\hat\varphi^a = \varphi^a + \Omega^a \hat t, \qquad \hat t = \lambda t,
\end{equation}
where $\Omega^a$ is the angular velocity with respect the coordinates $\varphi^a$, and $\lambda$ is a scaling factor.  The exponential factor becomes
\begin{equation}
\mathrm{e}^{- i \omega \hat{t} + i m_a \hat{\varphi}^a} = \mathrm{e}^{- i n_t t + i n_a \varphi^a}
\end{equation}
where
\begin{equation}
n_a = m_a, \qquad n_t = \lambda (\omega - \Omega^a m_a).
\end{equation}
The vacuum state is expected, after tracing over the states inside the horizon, to have a Boltzmann weighting factor as
\begin{equation}
\mathrm{e}^{- \hbar \frac{\omega - \Omega_0^a m_a}{T_H}} = \mathrm{e}^{- \frac{n_t}{T^t} - \frac{n_a}{T^a}},
\end{equation}
where $T_H$ is the Hawking temperature and $\Omega_0^a$ are angular velocities at the extremal horizon.  Computing the Frolov-Thorne temperatures is  straightforward
\begin{equation}
T^t = \frac{\lambda T_H}{\hbar}, \qquad T^a = - \frac{T_H}{\hbar (\Omega^a - \Omega_0^a)}.
\end{equation}
In dealing with extremal black holes, $T_H \to 0, \; \Omega^a \to \Omega_0^a$ and the ratio, which is finite, defines the Frolov-Thorne temperature
\begin{equation}
T_{FT}^a = - \lim_{r \to r_0} \frac{T_H}{\hbar (\Omega^a - \Omega_0^a)}.
\end{equation}

\subsection{Entropy}
%%%%%%%%%%%%%%%%%%%%%%%%%%%%%%%%%%%%%%%%%%%%%%%%%%%%%%%%%%%%%%%%%%%%%%
The Bekenstein-Hawking entropy can be calculated for the black hole.  As a check of the duality of the black hole and the conformal field theory, the CFT entropy can be computed via the Cardy formula (with no summation over the indices)
\begin{equation}
S_{CFT} = \frac{\pi^2}3 c^a T_{FT}^a.
\end{equation}
In all cases checked so far where this approach has been used, the entropies of the black hole and CFT match.

%%%%%%%%%%%%%%%%%%%%%%%%%%%%%%%%%%%%%%%%%%%%%%%%%%%%%%%%%%%%%%%%%%%%%%
\section{Black holes in 5D minimal supergravity}
%%%%%%%%%%%%%%%%%%%%%%%%%%%%%%%%%%%%%%%%%%%%%%%%%%%%%%%%%%%%%%%%%%%%%%

In the paper, we focus on the general rotating charged black holes of the five-dimensional minimal supergravity action
\begin{equation}
S_5 = \frac1{16 \pi G_5} \left[ \int d^5x \sqrt{- \hat g} \left( \hat R - \hat F^2 \right) - \frac8{3 \sqrt3} \int \hat F \wedge \hat F \wedge \hat A \right],
\end{equation}
where $\hat F = d \hat A$. The equations of motion are
\begin{equation} \label{EOM}
\hat G_{\mu\nu} = 2 \left( \hat F_{\mu\alpha} \hat F_\nu{}^\alpha - \frac14 \hat g_{\mu\nu} \hat F^2 \right), \qquad \hat D_\mu \hat F^{\mu\nu} = \frac1{2 \sqrt3 \sqrt{-\hat g}} \, \epsilon^{\nu\alpha\beta\gamma\delta} \hat F_{\alpha\beta} \hat F_{\gamma\delta},
\end{equation}
and the solution is supersymmetric if it admits a constant spinor to the Killing spinor equation
\begin{equation} \label{SUSYeq}
\left[ d + \frac14 \omega_{ab} \Gamma^{ab} + \frac{i}{4 \sqrt3} \left( \mathrm{e}^a \Gamma^{bc}{}_a F_{bc} - 4 \mathrm{e}^a \Gamma^b F_{ab} \right) \right] \eta = 0.
\end{equation}
Readers interested in more details can refer to \cite{Gauntlett:2002nw, Gauntlett:1998fz, Emparan:2008eg}.

\subsection{General solution}
%%%%%%%%%%%%%%%%%%%%%%%%%%%%%%%%%%%%%%%%%%%%%%%%%%%%%%%%%%%%%%%%%%%%%%
The general charged solutions of (\ref{EOM}) with two angular momenta have been obtained in \cite{Chong:2005hr} with the metric components, in $(t, r, \theta, \phi, \psi)$ coordinates\footnote{The ranges of angular coordinates are $0 \le \theta < \pi/2, \; 0 \le \phi, \psi < 2 \pi$.},
\begin{eqnarray}
g_{tt} &=& - \left( 1 - \frac{2 \mu}{\rho^2} + \frac{q^2}{\rho^4} \right),
\nonumber\\
g_{t\phi} &=& - \frac{a (2 \mu \rho^2 - q^2) + b q \rho^2}{\rho^4} \sin^2\theta,
\nonumber\\
g_{t\psi} &=& - \frac{b (2 \mu \rho^2 - q^2) + a q \rho^2}{\rho^4} \cos^2\theta,
\nonumber\\
g_{\phi\phi} &=& (r^2 + a^2) \sin^2\theta + \frac{a^2 (2 \mu \rho^2 - q^2) + 2 a b q \rho^2}{\rho^4} \sin^4\theta,
\nonumber\\
g_{\psi\psi} &=& (r^2 + b^2) \cos^2\theta + \frac{b^2 (2 \mu \rho^2 - q^2) + 2 a b q \rho^2}{\rho^4} \cos^4\theta,
\nonumber\\
g_{\phi\psi} &=& \frac{a b (2 \mu \rho^2 - q^2) + (a^2 + b^2) q \rho^2}{\rho^4} \sin^2\theta \cos^2\theta,
\nonumber\\
g_{rr} &=& \frac{\rho^2}{\Delta}, \qquad g_{\theta\theta} = \rho^2,
\end{eqnarray}
where
\begin{equation}
\rho^2 = r^2 + a^2 \cos^2\theta + b^2 \sin^2\theta, \qquad \Delta = \frac{(r^2 + a^2) (r^2 + b^2) + q^2 + 2 a b q}{r^2} - 2 \mu.
\end{equation}
The gauge potential is
\begin{equation}
A = \frac{\sqrt3 q}{2 \rho^2} (dt - a \sin^2\theta d\phi - b \cos^2\theta d\psi) .
\end{equation}

These solutions have four parameters $\mu$ (mass), $q$ (charge), $a, b$ (angular momenta) and the corresponding physical quantities are
\begin{equation}
M = \frac{3 \pi \mu}{4 G_5}, \qquad Q = \frac{\sqrt3 \pi q}{2 \sqrt{G_5}}, \qquad J^\phi = \frac{\pi}{4 G_5} (2 \mu a + b q), \qquad J^\psi = \frac{\pi}{4 G_5} (2 \mu b + a q).
\end{equation}
The angular velocities at the horizon are
\begin{equation}
\Omega_+^\phi = \frac{a (r_+^2 + b^2) + b q}{(r_+^2 + a^2) (r_+^2 + b^2) + a b q}, \qquad \Omega_+^\psi = \frac{b (r_+^2 + a^2) + a q}{(r_+^2 + a^2) (r_+^2 + b^2) + a b q},
\end{equation}
and the surface gravity and Bekenstein-Hawking entropy are
\begin{equation}
\kappa = \frac{r_+^4 - (a b + q)^2}{r_+ [(r_+^2 + a^2) (r_+^2 + b^2) + a b q]}, \qquad S_{BH} = \frac{\pi^2 [(r_+^2 + a^2) (r_+^2 + b^2) + a b q]}{2 G_5 \hbar r_+}.
\end{equation}
Here the subscript ``$+$'' denotes the outer horizon, the largest root of $\Delta(r) = 0$.

In the extremal limit (which generically is not supersymmetric)
\begin{equation}
q = \mu - \frac12 (a + b)^2 ,
\end{equation}
the horizons are degenerate and equal to
\begin{equation}
r_0^2 = \mu - \frac12 (a^2 + b^2) .
\end{equation}
Quantities associated to the extremal horizon will be labeled by the subscript ``$0$'' in what follows.

The Frolov-Thorne temperatures can be computed via
\begin{equation}
T_{FT} = - \frac1{\hbar} \lim_{r_+ \to r_0} \frac{\partial_{r_+} T_H}{\partial_{r_+} \Omega_+} = - \frac1{2 \pi} \lim_{r_+ \to r_0} \frac{\partial_{r_+} \kappa}{\partial_{r_+} \Omega_+},
\end{equation}
and the results are
\begin{eqnarray}
T_{FT}^\phi &=& \frac{(r_0^2 + a^2) (r_0^2 + b^2) + a b q}{\pi r_0 [a r_0^2 + b (2 \mu - a^2)]} ,
\\
T_{FT}^\psi &=& \frac{(r_0^2 + a^2) (r_0^2 + b^2) + a b q}{\pi r_0 [b r_0^2 + a (2 \mu - b^2)]} \ .
\end{eqnarray}

The near horizon geometry of the extremal solution is found by taking the limit
\begin{equation}
\phi \to \phi + \Omega_0^\phi \, t, \quad \psi \to \psi + \Omega_0^\psi \, t, \qquad r \to r_0 + \epsilon \, r, \qquad t \to \epsilon^{-1} \, \frac{\mu + \frac12 (a + b)^2}{2 r_0} \, t
\end{equation}
so the metric components become
\begin{eqnarray}
g_{tt} &=& - \frac{\mu^2 - 2 (a + b)^2 \mu + \frac14 (a + b)^3 (3a - 4a \cos^2\theta + 3 b - 4 b \sin^2\theta)}{\rho_0^4} r^2,
\nonumber\\
g_{t\phi} &=& \frac{r \sin^2\theta}{\rho_0^4} \Bigl[ (a + 2 b) \mu^2 - (a + b)^2 (2b - 3a \cos^2\theta - 3b \sin^2\theta) \mu
\nonumber\\
&& + \frac14 (a + b)^3 [ 2 (a - b)^2 \cos^2\theta \cos2\theta - a^2 - a b]  \Bigr] ,
\nonumber\\
g_{t\psi} &=& \frac{r \cos^2\theta}{\rho_0^4} \Bigl[ (2 a + b) \mu^2 - (a + b)^2 (2a - 3a \cos^2\theta - 3b \sin^2\theta) \mu
\nonumber\\
&& - \frac14 (a + b)^3 [ 2 (a - b)^2 \sin^2\theta \cos2\theta + b^2 + a b]  \Bigr] ,
\nonumber\\
g_{\phi\phi} &=& \left( \mu^2 + \frac{a^2 - b^2}2 \right) \sin^2\theta + \frac{a^2 (2 \mu \rho_0^2 - q^2) + 2 a b q \rho_0^2}{\rho_0^4} \sin^4\theta ,
\nonumber\\
g_{\psi\psi} &=& \left( \mu^2 + \frac{b^2 - a^2}2 \right) \cos^2\theta + \frac{b^2 (2 \mu \rho_0^2 - q^2) + 2 a b q \rho_0^2}{\rho_0^4} \cos^4\theta ,
\nonumber\\
g_{\phi\psi} &=& \frac{a b (2 \mu \rho_0^2 - q^2) + (a^2 + b^2) q \rho_0^2}{\rho_0^4} \sin^2\theta \cos^2\theta ,
\nonumber\\
g_{rr} &=& \frac{\rho_0^2}{4 r^2} , \qquad\qquad g_{\theta\theta} = \rho_0^2 ,
\end{eqnarray}
where
\begin{equation}
\rho_0^2 = \mu - \frac12 (a^2 - b^2) (1 - 2\cos^2\theta) = \mu + \frac12 (a^2 - b^2) \cos2\theta .
\end{equation}

For the following boundary condition, in coordinates order ($t, r, \theta, \phi, \psi$),
\begin{equation}
h_{\mu\nu} = \delta g_{\mu\nu} = \left( \begin{array}{ccccc} r^2 & r^{-2} & r^{-1} & r & r \\ & r^{-3} & r^{-2} & r^{-1} & r^{-1} \\ & & r^{-1} & r^{-1} & r^{-1} \\ & & & 1 & 1 \\ & & & & 1 \end{array} \right)
\end{equation}
the general diffeomorphism generators are
\begin{eqnarray}
\zeta &=& \left[ C + O(r^{-3}) \right] \partial_t + \left[ - r \epsilon_\phi'(\phi) - r \epsilon_\psi'(\psi) + O(1) \right] \partial_r 
\nonumber\\
&& + O(r^{-1}) \partial_\theta + \left[ \epsilon_\phi(\phi) + O(r^{-2}) \right] \partial_\phi + \left[ \epsilon_\psi(\psi) + O(r^{-2}) \right] \partial_\psi.
\end{eqnarray}
After taking mode expansion, $\epsilon_\phi(\phi) = - \exp(i n \phi), \epsilon_\psi(\psi) = - \exp(i n \psi)$ there are two sets of the Virasoro algebra generators
\begin{equation}
\zeta^\phi_n = - i n r \mathrm{e}^{- i n \phi} \partial_r - \mathrm{e}^{- i n \phi} \partial_\phi, \qquad \zeta^\psi_n = - i n r \mathrm{e}^{- i n \psi} \partial_r - \mathrm{e}^{- i n \psi} \partial_\psi,
\end{equation}
and we compute with Maple the associated charges (\ref{Cterm})
\begin{eqnarray}
Q^\phi_{mn} &=& - \frac{i \pi}{8 G_5} \Bigl( [a r_0^2 + b (2 \mu - a^2)] m^3 + \lambda^\phi m \Bigr) \delta_{m+n,0},
\\
Q^\psi_{mn} &=& - \frac{i \pi}{8 G_5} \Bigl( [b r_0^2 + a (2 \mu - b^2)] m^3 + \lambda^\psi m \Bigr) \delta_{m+n,0},
\end{eqnarray}
where
\begin{eqnarray}
\lambda^\phi &=& [2\mu + (a + b)^2] r_0^2 \int_0^{\pi/2} \frac{\sin^3\theta \cos\theta}{\rho_0^6} \Bigl( 4 (a + 2 b) \mu^2 + 4 (a + b)^2 [ b + 3 (a - b) \cos^2\theta ] \mu
\nonumber\\
&& \qquad - (a + b)^3 [ (a - b)^2 (1 + 2 \cos^2\theta - 4 \cos^4\theta) - b^2 + 3 a b ] \Bigr),
\nonumber\\
\lambda^\psi &=& [2\mu + (a + b)^2] r_0^2 \int_0^{\pi/2} \frac{\sin\theta \cos^3\theta}{\rho_0^6} \Bigl( 4 (2 a + b) \mu^2 + 4 (a + b)^2 [ a - 3 (a - b) \sin^2\theta ] \mu
\nonumber\\
&& \qquad - (a + b)^3 [ (a - b)^2 (1 + 2 \sin^2\theta - 4 \sin^4\theta) - a^2 + 3 a b ] \Bigr).
\end{eqnarray}
The explicit values of $\lambda^{(\phi,\psi)}$ are not relevant for determining the central charges as they can be redefined by a c-shift of the generators. Finally, the corresponding central charges can be read out from the relation (\ref{Ccharge})
\begin{eqnarray}
c^\phi &=& \frac{3 \pi}{2 G_5 \hbar} \left[ a r_0^2 + b (2 \mu - a^2) \right] = \frac{6}{\hbar} J^\psi,
\\
c^\psi &=& \frac{3 \pi}{2 G_5 \hbar} \left[ b r_0^2 + a (2 \mu - b^2) \right] = \frac{6}{\hbar} J^\phi.
\end{eqnarray}
It is an interesting fact that the central charges are proportional to the angular momentum as was the case for the extremal Kerr near horizon limit.  One unusual aspect however is that the central charges are not related to angular momentum in the same direction, $c^{\phi,\psi} \neq \frac{6}{\hbar} J^{\phi,\psi}$.  However it is precisely this property that allows there to be a consistent entropy calculation for the  conformal field theories associated to the two different angular directions as shown below.  Also because the angular momentum components form a vector, we will say that the central charges form a {\it central charge vector} $\vec{C}$.  Physically, angular momentum can be rotated without affecting the system.  The invariance of the central charge vector will be further discussed in Subsec.~C.

It is straightforward to check that the entropy of the CFT given by the Cardy matches the Bekenstein-Hawking entropy
\begin{equation}
S_{CFT} = S_{BH} = \frac{\pi^2 [(r_0^2 + a^2) (r_0^2 + b^2) + a b q]}{2 G_5 \hbar r_0} =\frac{\pi^2}3 c^\phi \, T_{FT}^\phi = \frac{\pi^2}3 c^\psi \, T_{FT}^\psi .
\end{equation}

Notice for example that for $a < 0$, the CFT associated with the angular coordinate $\psi$ has negative central charge and temperature.  This point will be further discussed in Subsec.~C.

\subsection{BMPV solution}
%%%%%%%%%%%%%%%%%%%%%%%%%%%%%%%%%%%%%%%%%%%%%%%%%%%%%%%%%%%%%%%%%%%%%%
Setting $b = - a = \omega$ and taking the extremal solution ($q = m$ in this case), we recover the Breckenridge-Myers-Peet-Vafa (BMPV) \cite{Breckenridge:1996is} black holes.  To obtain the standard form it is necessary to change coordinates $r^2 \to r^2 - \omega^2$, or equivalently use  $\rho$ as the new radial coordinate
\begin{eqnarray} \label{BMPV}
ds^2 &=& - \left( 1 - \frac{\mu}{r^2} \right)^2 \left( dt - \frac{\mu \omega \sin^2\theta}{r^2 - \mu} d\phi + \frac{\mu \omega \cos^2\theta}{r^2 - \mu} d\psi \right)^2 + \left( 1 - \frac{\mu}{r^2} \right)^{-2} dr^2 + r^2 d\Omega_3^2,
\nonumber\\
A &=& \frac{\sqrt3 \mu}{2 r^2} (dt + \omega \sin^2\theta d\phi - \omega \cos^2\theta d\psi) .
\end{eqnarray}
The three sphere metric $d\Omega_3^2 = d\theta^2 + \sin^2\theta d\phi^2 + \cos^2\theta d\psi^2$ and the ranges of angular coordinates are $0 \le \theta < \pi/2, \; 0 \le \phi, \psi < 2 \pi$.

The degenerate horizon is located at $r_0 = \sqrt\mu$ and the corresponding horizon area and angular velocities are
\begin{equation}
A = 2 \pi^2 \mu \sqrt{\mu - \omega^2}, \qquad \Omega_0^\phi = - \Omega_0^\psi = - \lim_{r \to r_0} \frac{\mu \omega (r^2 - \mu)}{r^6 - \mu^2 \omega^2} \to 0.
\end{equation}
The Hawking temperature of extremal black hole vanishes generically. However, its explicit form is essential to calculate the Frolov-Thorne temperature. We can start from the surface gravity of the non-extremal solution in \cite{Wu:2007gg} and take the extremal limit\footnote{The surface gravity of non-extremal solution is $\kappa = \frac{2 [ \mu r_+^2 - 2(\mu - q) \omega^2 - q^2]}{r_+^2 \sqrt{r_+^6 + 2 (\mu - q) \omega^2 r_+^2 - q^2 \omega^2}}$.}
\begin{equation}
\kappa = \lim_{r \to r_0} \frac{2 \mu (r^2 - \mu)}{r^2 \sqrt{r^6 - \mu^2 \omega^2}},
\end{equation}
which gives the Hawking temperature
\begin{equation}
T_H = \frac{\hbar}{2 \pi} \kappa = \lim_{r \to r_0} \frac{\hbar \mu (r^2 - \mu)}{\pi r^2 \sqrt{r^6 - \mu^2 \omega^2}} \to 0.
\end{equation}
It is now possible to calculate the Frolov-Thorne temperature associated with two angular coordinates $\phi$ and $\psi$
\begin{equation}
T_{FT}^\phi = - \frac{T_H}{\hbar \Omega_0^\phi} = \frac{\sqrt{\mu - \omega^2}}{\pi \omega}, \qquad T_{FT}^\psi = + \frac{T_H}{\hbar \Omega_0^\psi} = \frac{\sqrt{\mu - \omega^2}}{\pi \omega}.
\end{equation}
In the definition of the temperature associated to the $\psi$ direction we have added a negative sign due to orientation of the angular momentum.  Next, we compute the associated charges (\ref{Cterm}) of the near-horizon geometry of (\ref{BMPV}) which has an $SL(2,R) \times U(1)^2$ symmetry
\begin{eqnarray}
ds^2 &=& - r^2 \left( dt - \frac{\omega \sin^2\theta}{r} d\phi + \frac{\omega \cos^2\theta}{r} d\psi \right)^2 + \frac{\mu}{4 r^2} dr^2 + \mu d\Omega_3^2,
\end{eqnarray}
and the results are
\begin{eqnarray}
Q^\phi_{mn} &=& - \frac{i \pi}{8 G_5} \Bigl( \mu \omega \; m^3 + 2 (\mu - \omega^2) \omega \; m \Bigr) \delta_{m+n,0},
\\
Q^\psi_{mn} &=& - \frac{i \pi}{8 G_5} \Bigl( \mu \omega \; m^3 + 2 (\mu - \omega^2) \omega \; m \Bigr) \delta_{m+n,0}.
\end{eqnarray}
Again due to the orientation of the angular momentum we have added a negative sign into the charge along $\psi$.  Therefore, the corresponding central charges are
\begin{equation}
c^\phi = \frac{3 \pi}{2 G_5 \hbar} \mu \omega, \qquad c^\psi =  \frac{3 \pi}{2 G_5 \hbar} \mu \omega.
\end{equation}

Finally the entropy of the CFT can be computed with the Cardy formula  and the entropy is equal to the black hole Bekenstein-Hawking entropy
\begin{equation}
S_{CFT} = \frac{\pi^2}3 c^\phi \, T_{FT}^\phi = \frac{\pi^2}3 c^\psi \, T_{FT}^\psi = \frac{\pi^2}{2 G_5 \hbar} \mu \sqrt{\mu - \omega^2} = \frac{A}{4 G_5 \hbar}
\end{equation}

\subsection{Central charges as a vector}
%%%%%%%%%%%%%%%%%%%%%%%%%%%%%%%%%%%%%%%%%%%%%%%%%%%%%%%%%%%%%%%%%%%%%%
It is not clear how to interpret the fact that the entropy of the black hole can be reproduced by each of the two conformal field theories corresponding to the two different angular momenta.  Each central charge would at first sight correspond to its own conformal field theory.  Apparently, two different CFT descriptions associated with the angular coordinates, $\phi$ and $\psi$, and with different central charges describe the same gravity background.

We now turn to an additional and related property of the system.  The central charges $c^\phi$ and $c^\psi$ were earlier shown to be proportional to the angular momentum $J^\psi$ and $J^\phi$.  The angular momentum along the two directions can be rotated and transform as a vector.  In this subsection we perform this rotation explicitly through the Brown-Henneuax formalism and verify the transformation of the central charges.

In particular it is possible to redefine the two coordinates through linear combinations and rederive the central charges.  Consider the rotation between two angular coordinates
\begin{equation}
\phi = \cos\alpha \, \phi' + \sin\alpha \, \psi', \qquad \psi = \cos\alpha \, \psi' - \sin\alpha \, \phi',
\end{equation}
which changes the near horizon geometry to
\begin{eqnarray}
ds^2 = - r^2 \left[ dt - \frac{\omega (c_\alpha \sin^2\theta + s_\alpha \cos^2\theta)}{r} d\phi' + \frac{\omega (c_\alpha \cos^2\theta - s_\alpha \, \sin^2\theta)}{r} d\psi' \right]^2 + \frac{\mu}{4 r^2} dr^2 + \mu d\theta^2
\nonumber\\
+ \mu \left[ (c_\alpha^2 \sin^2\theta + s_\alpha^2 \cos^2\theta) d\phi'^2 + (c_\alpha^2 \cos^2\theta + s_\alpha^2 \sin^2\theta) d\psi'^2 + 2 s_\alpha c_\alpha (\sin^2\theta - \cos^2\theta) d\phi' d\psi' \right],
\nonumber
\end{eqnarray}
where we denoted $c_\alpha = \cos\alpha$ and $s_\alpha = \sin\alpha$.

Following the same approach as before, it is possible to calculate the conserved charges of (\ref{Cterm}) in the new coordinates
\begin{eqnarray}
Q^{\phi'}_{mn} &=& - \frac{i \pi}{8 G_5} (\cos\alpha + \sin\alpha) \Bigl( \mu \omega \; m^3 + 2 (\mu - \omega^2) \omega \; m \Bigr) \delta_{m+n,0},
\\
Q^{\psi'}_{mn} &=& - \frac{i \pi}{8 G_5} (\cos\alpha - \sin\alpha) \Bigl( \mu \omega \; m^3 + 2 (\mu - \omega^2) \omega \; m \Bigr) \delta_{m+n,0},
\end{eqnarray}
and the central charges become
\begin{equation}
c^{\phi'} = \frac{3 \pi}{2 G_5 \hbar} (\cos\alpha + \sin\alpha) \mu \omega, \qquad c^{\psi'} =  \frac{3 \pi}{2 G_5 \hbar} (\cos\alpha - \sin\alpha) \mu \omega.
\end{equation}

Therefore, in the two dimensional space of angular coordinates $(\phi, \psi)$, the central charges transform as a vector under rotation as should be expected since they are proportional to angular momentum
\begin{equation}
\left( \begin{array}{c} \phi' \\ \psi' \end{array} \right) = \left( \begin{array}{rr} \cos\alpha & - \sin\alpha \\ \sin\alpha & \cos\alpha \end{array} \right) \left( \begin{array}{c} \phi \\ \psi \end{array} \right) \quad \to \quad \left( \begin{array}{c} c^{\phi'} \\ c^{\psi'} \end{array} \right) = \left( \begin{array}{rr} \cos\alpha & - \sin\alpha \\ \sin\alpha & \cos\alpha \end{array} \right) \left( \begin{array}{c} c^\phi \\ c^\psi \end{array} \right).
\end{equation}

%We also can consider the rescaling transformation
%\begin{equation}
%(\phi', \; \psi') = (\beta_1 \, \phi, \; \beta_2 \, \psi),
%\end{equation}
%and the ranges of new coordinates should be $0 \le \phi' < 2 \pi \beta_1, \, 0 %\le \psi' < 2 \pi \beta_2$. Consequently we found the central charges are %transformed as
%\begin{equation}
%\left( c^{\phi'}, \; c^{\psi'} \right) = \left( \beta_1 \, c^\phi, \; \beta_2 \, %c^\psi \right).
%\end{equation}
%A final symmetry would be to interchange the angular momentum along the two %different directions.

The central charges associated to each angular coordinate is just a component of the ``central charge vector''.  If we take different values of the central charge vector $\vec{C}$ to correspond to different conformal field theories, it appears that there is a symmetry relating an infinite number of conformal field theory descriptions for each gravitational theory!  The choice of coordinates on the spacetime however should not be a physically relevant quantity, which suggests that these conformal field theories should be the same.  The conformal field theory should have a rotational and discrete symmetry.

We point out that the case $\alpha = \pi/4, \beta_1 = \beta_2 = \sqrt2$ corresponds to another convenient coordinates system for the BMPV black hole and its central charges have been discussed in \cite{Isono:2008kx}.

In the appendix, we discuss the central charge vector for the 5d Kerr string.

\subsection{Invariant central charge and a SUSY condition}
%%%%%%%%%%%%%%%%%%%%%%%%%%%%%%%%%%%%%%%%%%%%%%%%%%%%%%%%%%%%%%%%%%%%%%
Instead of dealing with a covariant central charge vector description, one way to obtain a coordinate invariant description of the system is to consider the invariant quantity
\begin{equation}
C^2 = (c^{\phi'})^2 + (c^{\psi'})^2 = \frac{36}{\hbar^2} J^2.
\end{equation}
This quantity, which is proportional to the square of the angular momentum, also has the interesting additional application.  Consider the central charges for the non-extremal two angular momentum solution
\begin{equation}
c^\phi = \frac{3 \pi}{2 G_5 \hbar} \left[ a r_0^2 + b (2 \mu - a^2) \right], \qquad c^\psi = \frac{3 \pi}{2 G_5 \hbar} \left[ b r_0^2 + a (2 \mu - b^2) \right].
\end{equation}
In general it is possible to find the supersymmetric solution $a = -b$ by directly solving the Killing spinor equations (\ref{SUSYeq}) from the gravitational side.  On the other hand one can ask what the equivalent condition would be from the conformal field theory perspective.  Given the fact that the CFT is described by its central charge, we propose that the SUSY condition is related to this new invariant central charge $C$.

\begin{figure}[ht]
\includegraphics[width=6cm, angle=-90]{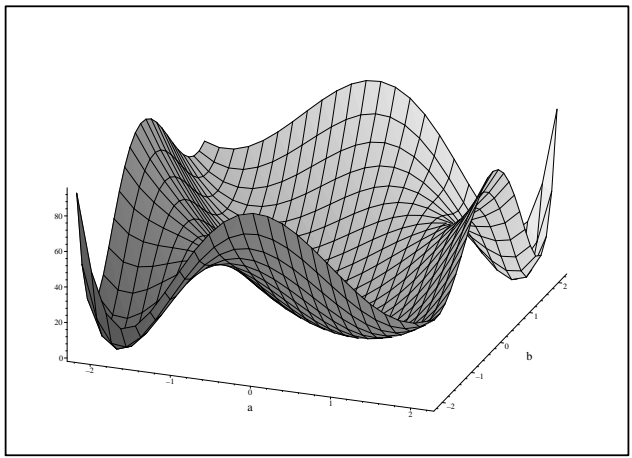}
\caption{The plot of the invariant central charge $C^2\propto J^2$ for the choice of parameter $\mu = 2$ and the normalization $\frac{3 \pi}{2 G_5 \hbar} = 1$.}  \label{InvCC}
\end{figure}

Supporting evidence for this comes from analyzing $C$ over the range of $a, b$ and noticing that the function features a valley along $a = - b$, see Fig.\ref{InvCC}.\footnote{From the figure, we can find there are several extremal points along the curve of $a = b$ on the surface. The points $a = b = 0, \; \pm \frac12 \sqrt{6\mu}$ correspond to local minima, while $a = b = \pm \frac12 \sqrt{2 \mu}$ are the saddle points.}  The valleys of the function are parallel to each other and despite the apparent complexity of the solution, they are all various parametrizations of the condition $J^\phi = -J^\psi$.

We next explore the question of how to define the SUSY condition which leads to $J^\phi = -J^\psi$.  One condition is that if we wish to minimize the central charge this would occur for $a=b=0$ which is the most supersymmetric solution.
\begin{equation}
\partial_a C=0, \qquad \partial_b C=0.
\end{equation}

We next explore several further conditions

Case I. Consider the vectors $\vec{C} = (c^\phi, c^\psi)$ and $\vec{T} = (T^\phi, T^\psi)$ and the invariant magnitudes $C = \sqrt{(c^\phi)^2 + (c^\psi)^2}$ and $T = \sqrt{(T^\phi)^2 + (T^\psi)^2}$.  The entropy can be written as $S = \frac{1}{2} (\frac{\pi^2}3 \vec{C}\cdot \vec{T})$.  Clearly several conditions are distinguished for the two vectors corresponding to their relative orientations $\cos\vartheta=\frac{\vec{C}\cdot \vec{T}}{C T}$.  If the vectors are orthogonal, $\vartheta=\pi/2$ then this leads to zero entropy.  The other distinguished cases in which the two vectors are parallel or anti-parallel, correspond to $\vartheta=0, \pi$.  Let use consider the values of $a, b$ where the vectors are parallel, $S = \frac12 (\frac{\pi^2}3 C T)$.  There are four solutions:
\begin{eqnarray}
b = a \quad &\to& \quad J^\phi = a (3 \mu - 2 a^2) = J^\psi,
\nonumber\\
b = - a \quad &\to& \quad J^\phi = \mu a = - J^\psi,
\nonumber\\
b = - a \pm \sqrt{6 \mu} \quad &\to& \quad J^\phi = 2 \mu (2 a \mp \sqrt{6 \mu}) = - J^\psi.
\end{eqnarray}
The first case is not supersymmetric while the second is the usual supersymmetric solution which reduces to the BMPV black hole. The last two cases appear to be an interesting new parametrization of the physical condition $J^\phi = -J^\psi$.  However, the corresponding square of the horizon radius, $r_0^2$, becomes negative, so these two cases do not give black hole solutions.  The condition that the charge and temperature vectors should be parallel while suggestive, is not sufficient to produce supersymmetry.

\begin{figure}[ht]
\includegraphics[width=6cm, angle=-90]{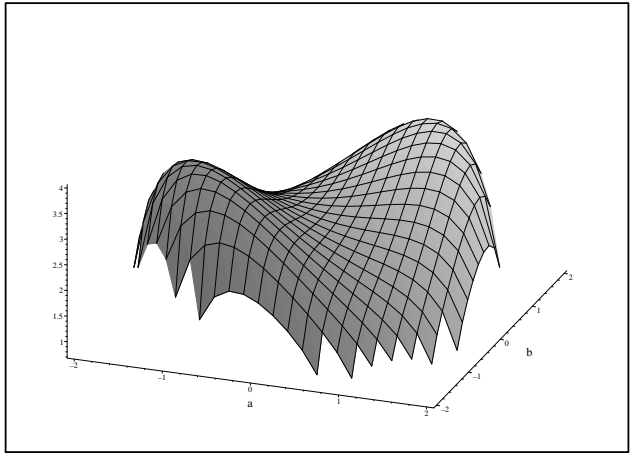}
\caption{The plot of the entropy $S$ for the choice of parameter $\mu = 2$ and the normalization $\frac{\pi^2}{2 G_5 \hbar} = 1$.}  \label{Entropy}
\end{figure}

Case II.  To find the supersymmetric state, we also consider finding the most ordered state and minimizing the entropy for fixed invariant $C$.   Finding the critical points of the entropy $\nabla S=0$ leads to the following solutions
\begin{eqnarray}
b = a &\to& S = \frac{\mu^2 + \mu a^2 - 2 a^4}{\sqrt{\mu - a^2}}, \quad S'' = - \frac{4\mu(\mu+2a^2)}{(3\mu - 2a^2) \sqrt{\mu - a^2}} < 0, \quad \mbox{(maximal)}
\nonumber\\
b = - a &\to& S = \mu \sqrt{\mu - a^2}, \quad S'' = \frac{12\mu}{\sqrt{\mu - a^2}} > 0, \quad \mbox{(minimal)}
\nonumber\\
b = - a + \sqrt{2 \mu} &\to& S = 2 \mu \sqrt{a} \sqrt{\sqrt{2 \mu} - a},
\nonumber\\
b = - a - \sqrt{2 \mu} &\to& S = i 2 \mu \sqrt{a} \sqrt{\sqrt{2 \mu} + a},
\nonumber\\
b = - a \pm i \sqrt{2 \mu} .
\end{eqnarray}
A plot of the entropy as a function of $a,b$ is given in Fig.~\ref{Entropy}.  To be precise we mention that $S''$ corresponds to making the entropy a function of one variable using the stated relations between $a,b$, and then minimizing with respect to the variable, $a$, under the condition that the central charge $C$ be constant.

Let us turn to an investigation of the solutions.  The solution $a=b$ does not minimize the entropy and corresponds to a local maximum.  The solution $a=-b$ is a local minimum and does lead to supersymmetry.  The third solution gives real entropy, but the second derivative is very complicated and its sign will depend on the values of parameters.  It is likely that this solution is not supersymmetric although its details have not been resolved.  The fourth solution can be discarded as it does not lead to real entropy.  The last two solutions lead to complex angular momentum.  It appears then that minimizing the entropy for fixed invariant charge may be a consistent relation for reproducing supersymmetric solutions.  The only apparent physical solution reproduces supersymmetry.

\begin{figure}[ht]
\includegraphics[width=6cm, angle=-90]{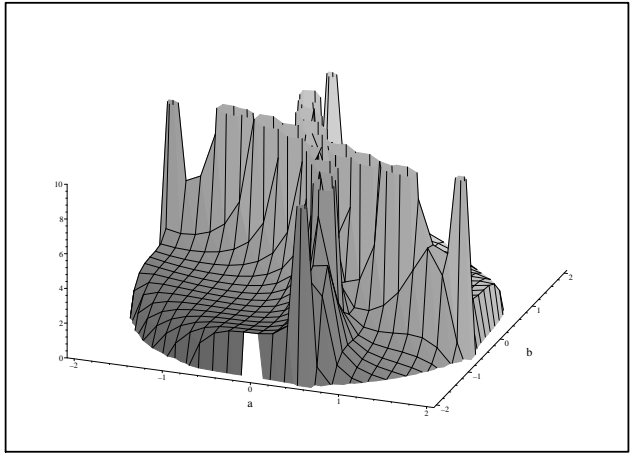}
\caption{The plot of the free energy $G$ for the choice of parameter $\mu = 2$ and the normalization $\frac{\pi}{2 G_5 \hbar} = 1$.}  \label{Fenergy}
\end{figure}

In addition to minimizing the entropy for fixed $C$, we also attempted to minimize the free energy $G = TS$,  for fixed central charge and zero enthalpy.  A plot of the free energy is given in Fig.~\ref{Fenergy}.  The solutions in this case exactly coincide with case II above.  The plot of the temperature is similar to the figure above for the free entropy.  Minimizing the temperature  for fixed $C$ also gives the same solutions as case II, providing further support for this supersymmetric condition.

It is also worth noting that in the supersymmetric limit the mass of the black hole and the angular momentum are proportional.

%%%%%%%%%%%%%%%%%%%%%%%%%%%%%%%%%%%%%%%%%%%%%%%%%%%%%%%%%%%%%%%%%%%%%%
\section{Conclusion}
%%%%%%%%%%%%%%%%%%%%%%%%%%%%%%%%%%%%%%%%%%%%%%%%%%%%%%%%%%%%%%%%%%%%%%

There are three main results in this paper.  The first is that the central charges and Virasoro generators of the rotating black holes in five-dimensional supergravity have been calculated, and the black hole Bekenstein-Hawking entropy matches the conformal field theory entropy.  This provides a dual holographic description of general rotating black holes in five dimensional minimal supergravity.  The second is that we have found that unlike in four dimensions, the five and higher dimensional black holes have a rotational symmetry which corresponds to an apparent change of the conformal field theory description.  We argue that this either suggests an large symmetry between conformal field theories or that dual description is physically described by a new central charge invariant.  Finally as an application, this central charge invariant is used to study the appearance of the supersymmetric BMPV limit of these black holes from the conformal field theory perspective.

It is still an open problem to more precisely describe the theory corresponding to the ``invariant charge'' $C^2 = \sum_a (c^{\varphi^a})^2=\frac{36}{\hbar^2} J^2$.  This charge seems useful and was  used to find the supersymmetric limit of the dual holographic description, although further examples are needed to confirm this result.  It would be interesting to work on a more detailed dual conformal field theory description for these general black holes.

\bigskip
{\center{\bf{Note added:}}}
Preprints \cite{Isono:2008kx} and \cite{Azeyanagi:2008dk} appeared while this article was near completion.  Both have overlapping discussion of the BMPV black holes and there is some overlap with the general non-supersymmetric black holes discussed here.

\section*{Acknowledgments}
%%%%%%%%%%%%%%%%%%%%%%%%%%%%%%%%%%%%%%%%%%%%%%%%%%%%%%%%%%%%%%%%%%%%%%
We are grateful to Wen-Yu Wen, Shuang-Qing Wu for their valuable comments. CMC was supported by the National Science Council of the R.O.C. under the grant NSC 96-2112-M-008-006-MY3 and in part by the National Center of Theoretical Sciences (NCTS) and the Center for Mathematics and Theoretic Physics at NCU.  JEW was supported in part by the Academic Center for Integrated Sciences at Niagara University, the New York State Academic Research and Technology Gen"NY"sis grant and the Niagara University Research Council.  JEW would also like to thank the University of Buffalo for support and valuable discussions.

%%%%%%%%%%%%%%%%%%%%%%%%%%%%%%%%%%%%%%%%%%%%%%%%%%%%%%%%%%%%%%%%%%%%%%
\appendix
\section{Invariance of charge and 5D Kerr String}
%%%%%%%%%%%%%%%%%%%%%%%%%%%%%%%%%%%%%%%%%%%%%%%%%%%%%%%%%%%%%%%%%%%%%%

In this appendix we check the vector property for the central charges for the five dimensional vacuum embedded Kerr black holes, namely the compactified Kerr strings\footnote{The ranges of angular coordinates are $0 \le \theta < \pi, \; 0 \le \phi, \, y < 2 \pi$.}
\begin{equation}
ds^2 = dy^2 - \frac{\Delta - a^2 \sin^2\theta}{\Sigma} \left[ dt + \frac{ 2 \mu a r \sin^2\theta}{\Delta - a^2 \sin^2\theta} d\phi \right]^2 + \Sigma \left( \frac{dr^2}{\Delta} + d\theta^2 + \frac{\Delta \sin^2\theta}{\Delta - a^2\sin^2\theta} d\phi^2 \right),
\end{equation}
where
\begin{equation}
\Sigma = r^2 + a^2 \cos^2\theta, \qquad \Delta = r^2 - 2 \mu r + a^2.
\end{equation}
The radii of outer and inner horizons are
\begin{equation}
r_\pm = \mu \pm \sqrt{\mu^2 - a^2},
\end{equation}
and the corresponding thermodynamical quantities are
\begin{equation}
T_H = \frac{\hbar \kappa}{2 \pi} = \frac{\hbar (r_+ - r_-)}{4\pi (r_+^2 + a^2)}, \qquad S_{BH} = \frac{A_+}{4 G_5 \hbar} = \frac{2 \pi^2 (r_+^2 + a^2)}{G_5 \hbar} = \frac{\pi (r_+^2 + a^2)}{G_4 \hbar},
\end{equation}
where $\kappa$ and $A_+$ are the surface gravity and area of the outer horizon and two Newton constants are related by $G_5 = 2 \pi G_4$.

The angular velocity near the extreme horizon, namely $a = \mu$ and at $r_0 = \mu$, is
\begin{equation}
\Omega_0 = \frac{2 \mu^2 r_0}{(r_0^2 + \mu^2)^2} = \frac1{2\mu},
\end{equation}
which we can remove by the coordinate transformation
\begin{equation}
\phi \to \phi + \Omega_0 \, t.
\end{equation}
After that one can take the following limit
\begin{equation}
r \to r_0 + \varepsilon r, \qquad t \to \varepsilon^{-1} (r_0^2 + \mu^2) t = 2 \mu^2 \varepsilon^{-1} \, t,
\end{equation}
and the near horizon solution is
\begin{equation}
ds^2 = dy^2 + \mu^2 (1 + \cos^2\theta) \left( - r^2 dt^2 + \frac{dr^2}{r^2} + d\theta^2 \right) + \frac{4 \mu^2 \sin^2\theta}{1 + \cos^2\theta} (d\phi + r dt)^2.
\end{equation}

Computing the charges
\begin{equation}
Q^\phi_{mn} = - \frac{i 2 \pi}{G_5} ( \mu^2 \; m^3 + 2 \mu^2 \; m ) \delta_{m+n,0}, \qquad Q^y_{mn} = 0,
\end{equation}
gives the central charges
\begin{equation}
c^\phi = \frac{24 \pi}{G_5 \hbar} \mu^2 = \frac{12 \pi}{G_4 \hbar} \mu^2, \qquad c^y = 0.
\end{equation}
The central charge $c^\phi$ is what remained for the dimensionally reduced four-dimensional Kerr black holes \cite{Guica:2008mu}.

By taking a rotation between coordinates ($\phi, y$)
\begin{equation}
\left( \begin{array}{c} \phi' \\ y' \end{array} \right) = \left( \begin{array}{rr} \cos\alpha & - \sin\alpha \\ \sin\alpha & \cos\alpha \end{array} \right) \left( \begin{array}{c} \phi \\ y \end{array} \right),
\end{equation}
we obtain the following associated charges
\begin{equation}
Q^\phi_{mn} = - \frac{i 2 \pi}{G_5} \mu^2 \cos\alpha (m^3 + 2 m) \delta_{m+n,0}, \qquad Q^y_{mn} = - \frac{i 2 \pi}{G_5} \mu^2 \sin\alpha (m^3 + 2 m) \delta_{m+n,0},
\end{equation}
and the following central charges
\begin{equation}
c^{\phi'} = \frac{24 \pi}{G_5 \hbar} \mu^2 \cos\alpha, \qquad c^{y'} = \frac{24 \pi}{G_5 \hbar} \mu^2 \sin\alpha.
\end{equation}
Again, the central charges transform like a vector
\begin{equation}
\left( \begin{array}{c} c^{\phi'} \\ c^{y'} \end{array} \right) = \left( \begin{array}{rr} \cos\alpha & - \sin\alpha \\ \sin\alpha & \cos\alpha \end{array} \right) \left( \begin{array}{c} c^\phi \\ c^y \end{array} \right)
\end{equation}
and the rescaling transformation also holds for the Kerr string.

%%%%%%%%%%%%%%%%%%%%%%%%%%%%%%%%%%%%%%%%%%%%%%%%%%%%%%%%%%%%%%%%%%%%%%

\end{document}